\documentclass[12pt,preprint]{aastex}

\newcommand{\etal}{{et al}\/.}

\begin{document}

\slugcomment{Draft of \today}

\shorttitle{Deep VLT $V$-band Imaging of a $z = 10$ Candidate}

\shortauthors{M. D. Lehnert \etal}

\title{Deep VLT $V$-band Imaging of the Field of a $z=10$ Candidate Galaxy:
Below the Lyman Limit?\altaffilmark{1}}

\altaffiltext{1}{Based on Director's Discretionary time observations
collected at ESO-VLT under program 273.A-5028(A)}

\author{M. D. Lehnert, N. M. F\"{o}rster Schreiber}
\affil{Max-Planck-Institut f\"ur extraterrestrische Physik,
Giessenbachstra\ss e, 85748 Garching bei M\"{u}nchen, Germany}
\and
\author{M. N. Bremer}
\affil{Department of Physics, University of Bristol, Tyndall Avenue,
Bristol BS8 1TL, UK}

\begin{abstract}

We present a deep 16.8~ks $V$-band image of the field of a candidate
$z{=}10$ galaxy magnified by the foreground ($z{=}0.25$) cluster A1835.
The image was obtained with FORS1 on VLT-Kueyen to test whether the
$V$-band lies below the Lyman limit for this very high redshift candidate.
A detection would unambiguously rule out that the source is at $z{=}10$.
The $3\sigma$ detection limit of the image in the area of the $z = 10$
candidate is $V_{\rm AB} = 28.0~{\rm mag}$ in a 2\arcsec -diameter
aperture (about 3 times the seeing FWHM of 0\farcs 7). No source at
the position of the candidate galaxy is detected down to this limit.
Formally, this is consistent with the $V$-band probing below the Lyman
limit in the rest-frame of a $z = 10$ source.  However, given the
recent non-detection of the object in a deep $H$-band exposure with
NIRI on Gemini North down to $H_{\rm AB} = 26.0~{\rm mag}$ ($3\sigma$
in a 1\farcs 4 aperture) and concerns about the detection of the reported
associated emission line, it may be possible that this source is spurious.
We discuss several astrophysical possibilities to explain the puzzling
nature of this source and find none of them compelling.

\end{abstract} 

\maketitle
\section{Introduction}
\label{intro}

Motivated by the discovery of high redshift, $z \sim 6$ quasars with what
appeared to be Gunn-Peterson troughs \citep{becker01, djorgovski01}, many
research groups began to search for the sources responsible for reionization
\citep[see {\it e.g.,}][]{L03,bremer04a, S04a, S04b, bunker03, ajiki03,
rhoads03, bouwens04, hu04}.  The WMAP result of the surprising detection
of a large Thompson electron optical depth of $\tau = 0.17 \pm 0.04$
\citep{kogut03} and questions about whether the intergalactic opacity at
$z \approx 6$ is due to the neutral intergalactic medium (IGM) or to discrete
absorbers \citep[e.g.,][]{sc02} have led observers to try and push discovery
techniques into the near-infrared (NIR) and beyond the redshift of the most
distant Sloan quasars \citep{pello04, kneib04}.

To reconcile the possible Gunn-Peterson troughs observed in high redshift
quasar spectra and the WMAP results, the fact that the ionizing photon density
at high redshift appears to be declining \citep[e.g.,][]{L03,bunker03} and
the rapidly increasing density necessary to ionize the IGM at successively
higher redshifts \citep[e.g.,][]{madau99} suggest that the universe may
have had a complex reionization history.  Indeed, these arguments led some
researchers to propose complex models such as extended partial reionization
\citep[e.g,][]{madau03, ricotti04a} or twice reionization
\citep[e.g.,][]{cen03, ciardi03, wyithe03}.  In these complex scenarios,
the discovery of even one high redshift star-forming galaxy provides
powerful constraints on the sources of reionization and on how star
formation proceeded in the early universe \citep[e.g.,][]{ricotti04b}.

But discovering galaxies at redshifts beyond $z\approx 6$ becomes
increasingly challenging.  Rest-frame emission longwards of the Lyman
limit is redshifted to observed $\lambda > 7500$~\AA\ and the faintness
of the galaxies makes them extremely difficult to detect.  The essential
spectroscopic confirmation is hampered by the dramatic increase in the
density of telluric OH emission bands at $\lambda > 7500$~\AA.  There are
regions $100 - 200$~\AA\ wide that are relatively devoid of OH lines
and narrow-band surveys for high redshift sources within these windows
have been successful \citep[e.g.,][]{hu04}.  Outside of these windows,
even determining redshifts of color selected galaxies is generally very
difficult at $z \ga 6$.

One technique developed to overcome the difficulty of detecting the most
distant galaxies takes advantage of gravitational lensing by an intervening
galaxy cluster to boost the apparent brightness of background sources.
This boost can be as much as a factor of $10 - 100$ along the critical lines
for lensing.  \cite{santos04} proved the feasibility of this technique
out to $z = 5.6$ and \cite{kneib04} discovered a probable lensed $z \sim 7$
Lyman break galaxy behind A2218.  The efficiency of gravitational lensing
assisted searches compared to blank field searches is, however, sensitive
to the slope of the luminosity function and thus its overall utility in
finding large numbers of high redshift galaxies has yet to be assessed.

\cite{pello04} reported the identification of a highly magnified
galaxy lying on a critical line of the $z = 0.25$ cluster A1835 (which
they denoted A1835-1916).  Their data set included broad-band optical
imaging from the {\em HST\/} and the CFHT along with NIR imaging and
spectroscopy with ISAAC at the VLT.  The object was undetected in the
$V,R,I$ optical bands, and only detected at $4\sigma$ in $H$ and $3\sigma$
in $K$.  The $J$-band detection quoted by \citet{pello04} is formally an
upper limit.  The optical non-detection, the large break between the $J$
and $H$ bands, and the blue $H - K_{\rm s}$ color ($H_{\rm AB} - K_{\rm
s,AB} < 0$) found by \citeauthor{pello04}could possibly indicate a young
galaxy at extremely high redshift.  In their $J$-band spectroscopy
\citeauthor{pello04}reported an emission line at 1.33745~\micron,
detected in two separate wavelength settings of the spectrograph and with
a combined significance of $4 - 8\,\sigma$.  The photometry, together
with the lensing model suggesting the source lies on a caustic for very
high redshift (with a magnification factor between 25 and 100 as being
most likely), led \citeauthor{pello04}to argue that the line is most
likely Ly$\alpha$ at $z = 10.0175$.  Unfortunately, the signal-to-noise
ratio of the spectrum is insufficient to show the telltale signature
of the line profile asymmetry of highly redshifted Ly$\alpha$ to rule
out other line identifications.  Given the uncertainties in the lensing
model, the most important piece of evidence upon which the conclusion
of $z = 10$ rests is the shape of the spectral energy distribution (SED)
measured from the imaging data.

However, the high redshift nature of this source has been recently
questioned.  Based on new independent $H$-band data obtained with NIRI at
the Gemini North telescope, \citet{bremer04b} did not detect the $z = 10$
candidate down to the $3\sigma$ limit of $H_{\rm AB} = 26.0~{\rm mag}$.
This limit is 1 magnitude deeper than the $4\sigma$ detection quoted by
\citet{pello04} in their ISAAC data.  This significantly weakens the main
evidence for a redshift of 10, which relies on the strength of the break
between the optical and $J$ bands and the $H$ band.  The photometry no
longer constrains the redshift and other identifications for the emission
line from a lower redshift source remain possible.

Before the \cite{bremer04b} results were obtained, we were awarded
Director's Discretionary Time with FORS1 at the VLT to conduct
$V$-band imaging of the field around the $z = 10$ candidate and push
the sensitivity to fainter levels than presented by \citet{pello04}.
A $V$-band detection would be decisive: it would demonstrate beyond
any doubt that the source is {\em not\/} at $z = 10$.  Besides probing
rest-frame wavelengths $\sim 500$~\AA\ for $z = 10$, well below the Lyman
limit, we chose the $V$-band to reach within a reasonable observing
time very sensitive limits compared to other optical or NIR bands and
thus provide the best chance of detecting a reddened $z = 1-2$ object,
which seems a likely alternative \citep[see][]{bremer04b}.

\section{Data and analysis}

\subsection{Observations and Data Reduction}

The $V$-band observations of A1835 were carried out during Director's
Discretionary Time on the nights of 2004 July 9, 16, and 19 (UT).  We used
FORS1 on the VLT-Kueyen telescope in imaging mode with a projected scale
of $\rm 0\farcs 200~pixel^{-1}$.  Forty-eight separate frames of 350~s
each were taken, for a total integration time of 16.8~ks.  Half of the
integration time was taken on 2004 July 16 and one-third on 2004 July
19, with only a small fraction (1/6) on 2004 July 9.  The conditions
were clear and the seeing was typical for the VLT in the optical,
varying from 0\farcs 5 to 1\farcs 0 with a median of about 0\farcs 7.
Each individual dithered exposure had a unique (non-redundant) pointing
position, centered around the location of the $z = 10$ candidate.

We reduced the data within IRAF as follows.  We subtracted the bias
from each frame using bias frames taken at the beginning and end of
each night.  We flattened each frame initially with flat field images
produced by a combination of dithered sky images taken during twilight.
From these bias-subtracted and flat-fielded images, we created a master
sky flat by combining all of the dithered exposures of the target
field, masking out all sources above $1\sigma$ of the average rms of
each frame (the rms calculated including the whole frame).  This flat
was divided into each dithered frame.  From the master sky flat, we
also generated the bad pixel map used to exclude the bad pixels in each
frame during the final combination.  We determined the offsets between
individual exposures by fitting the position of about 30 bright,
unsaturated stars.  We co-averaged the reduced, registered frames using
three different weightings: \footnote{The weight $w$ is applied
as multiplicative factor $w^{-2}$.} (1) uniform weighting, (2) weighting
by the seeing FWHM of each frame, which gives most weight to the frames
with best seeing, and (3) weighting by the FWHM times the squared root
of the average sky counts in each frame, which accounts for the rms
noise within the seeing disk and optimizes for point-source detection.
There is little variation in the sensitivity of the final images between
the three different schemes.  The latter two produce final combinations
with the best FWHM of $\sim 0\farcs 7$.  For the analysis, we adopted
the image combined following the third weighting scheme.

To flux-calibrate the data we measured the photometric zero points using
exposures of standard stars taken at the beginning and end of each night.
We compared these measurements to zero points estimated throughout
the month encompassing the range of observing dates.  All zero points
were in excellent agreement (within $\rm \lesssim0.02~mag$) and the final
calibration is based on the average measured by ESO during 2004 July when
our data were taken.  \footnote{The ESO FORS1 zero points are available at
http://www.eso.org.} The average sky brightness in the final image is $V =
21.5~{\rm mag~arcsec^{-2}}$, and varies among the individual frames by a
few tenths of a magnitude.  \footnote{We refer to the Vega photometric
system when not explicitely indicating AB magnitudes.  For the $V$
band, the AB correction is very small, with $V_{\rm AB} - V_{\rm Vega}
= 0.014~{\rm mag}$ for FORS1 accounting for the full system transmission.}

\subsection{Photometry}

\begin{figure}
\vspace{2cm}
\begin{center}
\includegraphics[scale=0.8]{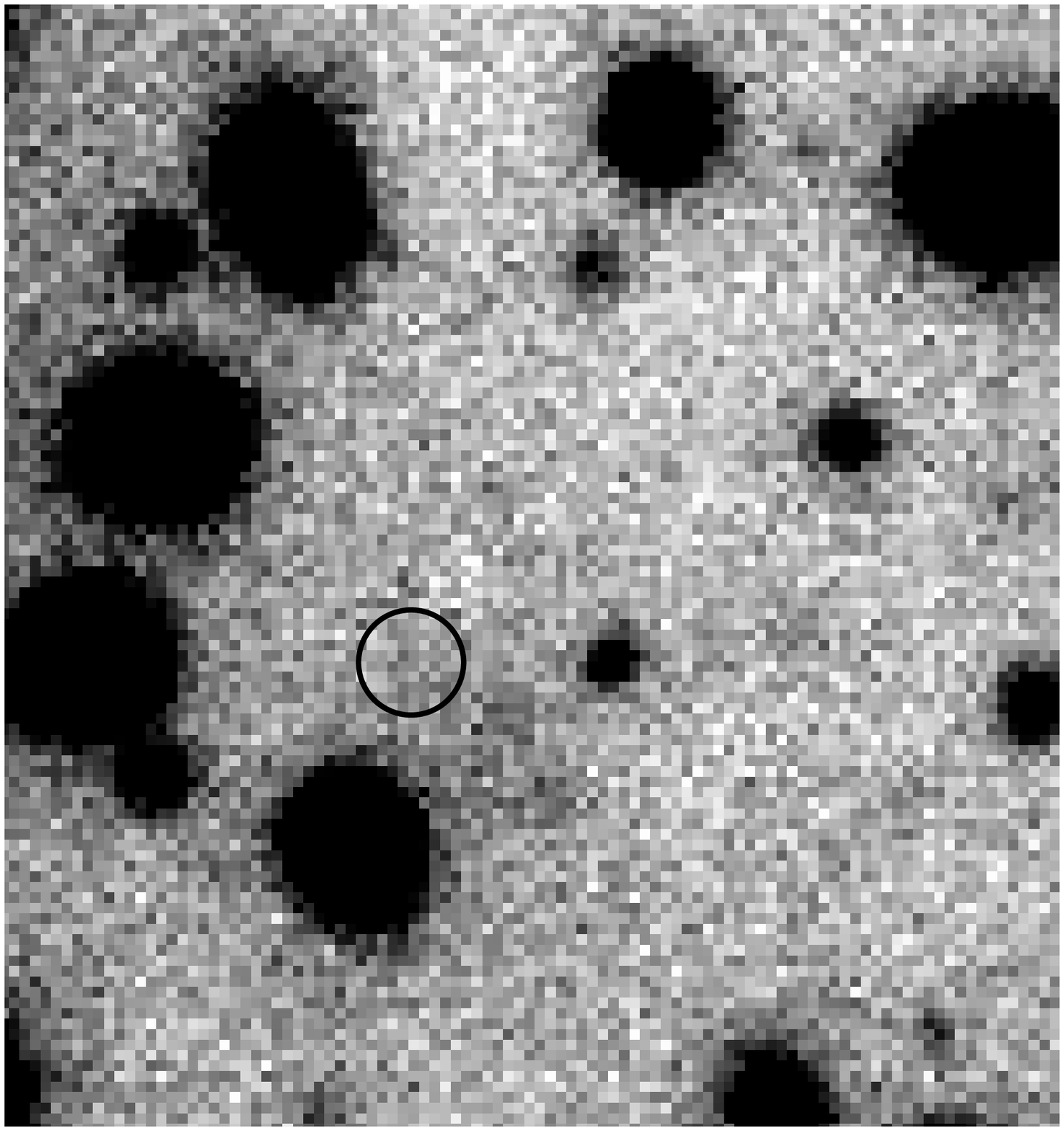}
\end{center}
\caption{Deep $V$-band FORS1 image of the region around the $z{=}10$
candidate object reported by \cite{pello04}.  The image is approximately
$20^{\prime\prime} \times 20^{\prime\prime}$ with seeing of $\sim 0\farcs
7$.  The circle indicates the location of the candidate object, and is 2.0
arcsec in diameter.  There is no evidence for emission from the candidate
in the V-band down to a 3$\sigma$-limit of $V_{\rm AB}=28.0~{\rm mag}$.
\label{fig-Vmap}}
\end{figure}

Figure~\ref{fig-Vmap} shows the $20^{\prime\prime} \times
20^{\prime\prime}$ region in our $V$-band image around the reported
location of the candidate $z = 10$ galaxy.  We do not detect any source
at this position down to the faintest levels reached in our data.
This implies a $3\sigma$ limit of $V_{\rm AB} > 28.0~{\rm mag}$ in a
2\arcsec -diameter aperture, as described below.  \footnote{This aperture
size corresponds to 3 times the seeing FWHM in the combined data and has
a negligible aperture correction except for the brightest point sources
and the most extended objects.  We note that it leads to conservative
limits for faint point sources, for which the S/N is optimized with an
aperture of 1.5 times the FWHM.} Our limit is 0.6~mag deeper than the
$3\sigma$ limit of \citet{pello04} given for a 0\farcs 6 aperture, or
about 3.5~mag deeper for an aperture of 3 times the FWHM (0\farcs 76)
of their $V$-band data, assuming uncorrelated Gaussian noise.

We assessed our ability to reliably set a meaningful upper limit to
any possible source at the position of the $z = 10$ candidate using
three different techniques.  First we established the pixel-to-pixel rms
variations in the background across the entire image.  For this purpose,
we used the Sextractor software \citep{bertin96}.  Sextractor estimates
the local background level in a mesh grid over the image.  These local
background estimates are iteratively clipped until they converge to
$\pm 3\sigma$ around the median of all meshes.  The histogram of all
remaining unclipped pixels is then used to determine the rms variations
of the background noise in the image.  This procedure yielded an
overall $3\sigma$ detection limit of $V_{\rm AB} = 27.6~{\rm mag}$
in a 2\arcsec -diameter aperture.  However, this limit is pessimistic
because our image is very crowded with cluster and background galaxies
at the depths reached: the surface density of faint sources is such
that there is typically only about 4\arcsec\ between adjacent objects.
Thus, the rms variations derived in this way partly reflect extended
light profiles of the myriad of galaxies in the image.

To obtain more realistic estimates of our detection limits in the area
around the $z = 10$ candidate, we focussed our analysis to the region
within $\pm 20$\arcsec\ of its reported location.  We placed 30-60
apertures 2\arcsec\ in diameter all away from the outer light profile of
any obvious object in this $\rm 1600~arcsec^{2}$ area.  This procedure was
repeated several times with differing numbers of apertures and positions
to ensure that our final estimates were not critically dependent on the
number of apertures or their exact placement. The average pixel-to-pixel
rms within these apertures implies a $3\sigma$ detection limit of
$V_{\rm AB} = 28.0~{\rm mag}$, with variations of $\approx 0.1~{\rm
mag}$ between the apertures.  The variation in the total residual flux
within each aperture implies a $3\sigma$ detection limit of $V_{\rm AB}
= 27.8~{\rm mag}$.

To further determine the robustness of our detection limits, we placed
20 artificial point sources each at 3 different AB magnitudes, 27.6,
27.8 and 28.0, within $\pm 20$\arcsec\ of the reported location of the
$z = 10$ candidate and avoiding all obvious sources in this region.
The high degree of crowding limits the number of point souces that
can be placed non-redundantly close to the position of the $z = 10$
candidate to about 20.  We recovered the sources down to $V_{\rm AB}
= 28.0~{\rm mag}$ at a rate similar to that expected for a $3\sigma$
detection limit (i.e. 50\%), and the dispersion between the measured and
true input brightness was $\approx \pm 0.3-0.4~{\rm mag}$.  Given the
excellent agreement between the two most robust methods, we adopted
$V_{\rm AB} = 28.0~{\rm mag}$ as the detection limit in a 2\arcsec\
aperture at the position of the $z = 10$ candidate.

\section{Discussion}
\label{discuss}

The non-detection in our VLT FORS1 $V-$band data down to a faint limit has
implications for the nature of the source investigated by \citet{pello04}.
Formally, a non-detection is consistent with the candidate having a
redshift of 10.  The Bessel $V$ band filter has an effective wavelength
of 5540~\AA\ and half maximum transmission at about 5000 and 6000~\AA.
For $z = 10$, these wavelengths correspond to 500~\AA\ and the range
$\approx 450 - 550$~\AA\ in the rest-frame --- well below the Lyman limit
and completely opaque.  The lower limit to the redshift, if the $V$-band
non-detection is caused solely by the IGM opacity below the Lyman limit,
is $\sim5.6$.  However, there is substantial opacity within the Lyman
forest at these redshifts and thus $V$-band Lyman break or drop-out
galaxies have redshifts that are usually less than this \citep[see
e.g.,][]{bremer04a}.

Although our $V$-band non-detection may allow for $z \ga 5.6$, or even $z =
10$, this is not the only interpretation possible in view of other lines
of evidence that have recently come to light.  There are certainly two
other hypotheses that are equally plausible, and perhaps significantly
more likely.

The strongest evidence for the $z = 10$ interpretation presented by
\citet{pello04} relied on the large break among the optical, $J$,
and $H$ bands, and the subsequent detection of an emission line
at 1.33745~\micron.  However, this evidence has now been called into
question.  \cite{bremer04b} did not detect the candidate object in their
new independent and $\rm 1~mag$ deeper $H$-band image from NIRI down
to $H_{\rm AB} = 26.0$ ($3\sigma$).  This greatly weakens the argument
based on the break strength (as well as the blue $H - K_{\rm s}$ color)
that supported the claim for the emission line being Ly$\alpha$ at $z
\approx 10$.  Furthermore, \cite{Weatherley04} have recently questioned
the robustness of the line detection.  However, this was based on a
re-analysis of the spectroscopic data used by \cite{pello04} and subtle
differences in reduction techniques could lead to contentious results.
It is probably reasonable to conclude that the significance of the line
is difficult to judge.

If the object is not at a redshift of 10, then what could it be?  One way
forward is to assume the reality of the line, use the sensitive $H$-band
limit of \cite{bremer04b}, and assume an alternative line identification
to determine if the emission properties are reasonable for a galaxy at
a lower redshift.  \citeauthor{bremer04b}argued that the source could
be a dwarf/HII galaxy at intermediate redshift and that lines such as
the [\ion{O}{2}]$\lambda\lambda 3726,3729$ doublet, [\ion{O}{3}]$\lambda
5007$, or H$\alpha$ at $1.0 < z < 2.6$ are the most likely alternatives to
Ly$\alpha$.  If it were [\ion{O}{3}], then the line width and luminosity
would be consistent with the local relationship for dwarf/\ion{H}{2}
galaxies between these quantities \citep[e.g.,][]{melnick00}.  In light
of our new $V$-band results, we wish to further this argument.

Assuming the line is [\ion{O}{3}] at $z = 1.67$ and adopting reasonable
cosmological parameters ($H_{0} = 70 {\rm km\,s^{-1}\,Mpc^{-1}}$,
$\Omega_{\rm m} = 0.3$, and $\Omega_{\Lambda} = 0.7$), our $V$-band limit
implies an unlensed absolute magnitude fainter than $-16.4~{\rm mag}$
at 2080~\AA.  Since there is no published detailed analysis of the mass
model of the intervening cluster, it is difficult to quantitatively
assess the magnification factor for $z = 1.67$ (\citeauthor{pello04}
suggested a magnification factor of between 25 and 100 for a source
at $z = 10$).  Conservatively, we assume no lensing; lensing will
only make the source intrinsically fainter.  The co-moving density
of galaxies in the local universe with absolute magnitude $-16.5$ at
2000~\AA\ is $\phi_{2000} \approx 10^{-1.5}~{\rm Mpc^{-3}\,mag^{-1}}$
\citep{treyer98}.  The number of such sources on the sky with $1.45 <
z < 1.95$ is then $\approx 50~{\rm arcmin^{-2}}$ if the UV luminosity
function does not evolve out to $z = 1.7$.  The luminosity function of
$z \approx 3$ Lyman break galaxies implies a roughly similar surface
density for such faint objects \citep{SLBGs4}.  Comparable densities
would be inferred to a factor of $\sim$1.5 if the line were H$\alpha$
at $z = 1.04$ or [\ion{O}{2}] at $z = 2.59$.

The surface density of sources down to the detection limit of our image
is $>$50 arcmin$^{-2}$ (modulo the incompleteness).  Therefore the
probability of finding and perhaps even mistaking it for a very high
redshift source is consequently high.  The immediate explanation is
that the source might simply be fainter.  However, this poses problems
when trying to reconcile the broad-band flux limits with the reported
emission line flux.

The $3\sigma$ $H$-band limit obtained by \citet{bremer04b} coupled with
the line flux given by \citet{pello04} implies an observed equivalent
width (EW) of $W_{\rm obs} > 170$~\AA\ (assuming a flat $f_{\nu}$
continuum between 1.65 and 1.34~\micron; if the source were redder
[bluer] as suggested by \citeauthor{bremer04b}, the EW limit would
be higher [lower]).  Assuming the line is [\ion{O}{3}] at $z = 1.67$,
the corresponding rest-frame EW is $W_{\rm rest} \ga 60$~\AA.  For the
alternative case of H$\alpha$ at $z = 1.04$, $W_{\rm rest} \ga 80$~\AA,
and of [\ion{O}{2}], $W_{\rm rest} \ga 45$~\AA.  Locally, such large
EWs are only found in galaxies with low metallicity and blue colors
\citep[][]{sullivan00}.

Although the \citeauthor{bremer04b}$H$-band limit sets a reasonably
stringent lower limit on the EW of the line, our $V$-band limit may
provide an even stronger constraint.  Locally, sources with large
emission line EWs are very blue, with ${\rm UV} - B$ colors always bluer
than for a flat $f_{\nu}$ spectrum \citep{sullivan00}.  For sources
at $z\approx1.7$ this implies observed $V_{AB}-H_{AB}\lesssim 0$.
Supposing again the source is at $z = 1.67$ and assuming conservatively
it has a flat $f_{\nu}$ spectrum, our limit of $V_{\rm AB} > 28.0~{\rm
mag}$ implies an $H$-band limit of $H_{\rm AB} > 28.0~{\rm mag}$ and,
repeating the reasoning above, a rest-frame EW for [\ion{O}{3}] of
$W_{\rm rest} > 400$~\AA ($\gtrsim$500~\AA\ if H$\alpha$, $\gtrsim$300~\AA\
if [\ion{O}{2}]).  For a bluer spectrum as observed locally, the line
EW limit only increases.  A similar argument would apply to any other
optical line.  Such high EWs are rarely observed, would require a very
young age ($\rm < 10~Myr$) and perhaps an initial mass function heavily
skewed toward massive stars.

Summarizing the above arguments, our sensitive $V$-band limit
implies a faint intrinsic UV ($\sim 2000$~\AA) absolute magnitude.
Plausible luminosity functions predict a high surface density of
such objects and thus a high probability of detection if the source
were just bright enough to match the $3\sigma$ limiting magnitude of
our $V$-band image.  No object is detected at the expected location
despite the high density $\ga 50~{\rm arcmin^{-2}}$ of $3 - 4\sigma$
sources in the data.  Postulating then that the object could be fainter,
the lower limits on the rest-frame EW of the emission line become very
large and the probability of detecting such a source becomes very low.
Thus, in view of the simple astrophysics of line emission, the scenario
of an intermediate redshift dwarf does not appear very likely given
the difficulties in reconciling the broad-band limits with the emission
line flux.  Additionally, although largely circumstantial, the arguments
could perhaps cast doubt on the reported line flux \citep{Weatherley04}.

The other possible scenario is that the source may be transient or
variable.  The NIR ISAAC images and spectra presented by \citet{pello04}
were taken in 2003 February and late June to early July, respectively.
The NIRI $H$-band data were collected during two nights in 2004 late
May and early June.  The $V$-band data discussed here were obtained in
2004 mid-July.  Overall, these observations span a period of about a
year and a half.

Given the ecliptic latitude of the A1835 field at about 14$^{\circ}$,
one hypothesis could be that the source is a solar system object with
a large proper motion.  The tightest constraint comes from the time
over which all of the \citet{pello04} ISAAC $H$-band data were taken.
The observations were carried out over 24 hours on 2003 February 15
and 16 and the seeing in the final $H$-band image is $\sim 0\farcs 5$.
Conservatively, for the source not to appear extended would imply
an angular velocity of $< 0\farcs 5 / 24~{\rm h}$ or $< 0\farcs 02
{\rm h}^{-1}$.  If the source is bound to the Sun and consistent with
a distant object like Pluto, for example, with a tangential velocity
of $5~{\rm km\,s^{-1}}$, it would lie at more than D=1240~AU.  To be
unresolved in the 0\farcs 5 $H$-band image at the distance derived
from the proper motion limit, the source would have to be smaller than
$\sim 10$ Jupiter diameters.  The brightness of an object illuminated
by reflected sunlight and assuming an albedo of one, is related to its
angular diameter, $\theta$, by m = m$_{\odot}$ - 5 log($\theta$/2D).
From the size calculated above the source would have $H\approx$55 mag.
An object with the magnitude as claimed by \cite{pello04} would have to
be a million times larger than what we roughly estimated and thus would
violate the size constraint by 6 orders of magnitude. It is possible to
relax these constraints by an order of magnitude, but even given more
optimistic estimates, a solar system object seems highly unlikely.

It is more difficult to estimate the probably that the source is a
high redshift supernova, $\gamma$-ray burst or perhaps something more
exotic \citep[e.g.,][and references therein]{stern04}.  \cite{dahlen04}
determined a SN rate of about 5.5 $\times$ 10$^{-4}$ SN yr$^{-1}$
Mpc$^{-3}$ at $<z>$$\approx$0.7-0.8 and using a model for the
star-formation rate density evolution, they estimate about a factor
of 2 increase in the SN rate at z$\approx$2.  The magnitude at which
\citeauthor{pello04}claim to have detected the source is $H_{AB}=25.0$.
It is difficult to know what the redshift of the source might be,
especially since the photometry for a transient provides no constraints
on the redshift.  It is worth noting that distant SNe generally have
optical magnitudes fainter than the \citeauthor{pello04}detection
\citep[e.g.,][]{stolger04,dahlen04} and given the relatively blue
SED of SNe this means that the $z{=}10$ candidate would have been one
of the brighter SNe (it is of course magnified by an unknown amount
which could account for its relative brightness).  At any rate, we can
estimate the chances of finding a SNe serendipitously in their field.
Assuming that the SN was caught within 30 days of its peak, that the
redshift range of the SN was 0.5-1.0 \citep[roughly consistent with the
range in][]{dahlen04}, and that the area in the detection image was 6
arcmin$^2$, we would have expected 0.22 SN in the field based on the
\cite{dahlen04} SNe rate density.  Thus the candidate as an intermediate
redshift SN cannot be ruled out.  However, this is surely optimistic given
the magnitude of the detection claimed in the \citeauthor{pello04}$H$-band
image and the unknown magnification.  Lensing would decrease the
probability by decreasing the effective area sampled.  Requiring the
SN to be observed more closely to its peak in order make detection more
likely would (linearly) lower the relative probability of observing it.

Neither of the transient hypotheses can explain the line detection.
One would have to postulate that it comes from the host galaxy in the case
of a SNe from which there has been no subsequent continuum detection.
Given the EW arguments we made previously, this would have to be a
very unusual object to escape detection.  For a solar system object,
the gap between the dates of the imaging and spectroscopy means that
the source would have moved a significant distance from the discovery
position and not within the subsequent slit spectroscopy.  Therefore,
none of the possible transient hypotheses explain simply or logically
all of the claims in \cite{pello04}.  Given the $H$-band non-detection
of the source down to faint limits by \citeauthor{bremer04b}and now in
the V-band, and, the unlikely nature of any galaxy with the properties
(lower-limits) observed, it is tempting to conclude that the most likely
explanation is that the source is spurious.

\acknowledgments

We would like thank the referee, Pat McCarthy, for his comments, and
Roser Pello and Daniel Schaerer for their arguments and insights.
We wish to thank the ESO DG, Catherine Cesarsky for the generous
allocation of observing time and Bruno Leibundgut for his curiosity
and support.  We are grateful to the ESO USG, DMD, and Paranal staff for
conducting these observations, Mario van den Ancker and Roberto Mignani,
in particular, for their dedication and effort.

\end{document}